\documentclass[aps,prd,nofootinbib,twocolumn]{revtex4}
\usepackage{amssymb}
\usepackage{amsmath,color}
\usepackage{epsfig}
\usepackage{graphicx,epsf,epsfig}
\usepackage{bbm}
\usepackage{subfigure}
\usepackage{multirow}
\usepackage{setspace}
\usepackage{verbatim}
\usepackage{epstopdf}
\usepackage{hyperref}
\usepackage{float}
\usepackage{ulem}
\usepackage[utf8]{inputenc}
\usepackage{color}

\definecolor{purple}{rgb}{0.87, 0.0, 1.0}

\newcommand{\be}{\begin{eqnarray}}
\newcommand{\ee}{\end{eqnarray}}

\begin{document}

\title{Do the observational data favor a local void?}

\author{Rong-Gen Cai$^{1,2,3}$}
\email{cairg@itp.ac.cn}

\author{Jia-Feng Ding$^{1,2}$}
\email{dingjiafeng@itp.ac.cn}

\author{Zong-Kuan Guo$^{1,2,3}$}
\email{guozk@itp.ac.cn}

\author{Shao-Jiang Wang$^{2,4}$}
\email{schwang@cosmos.phy.tufts.edu}

\author{Wang-Wei Yu$^{1,2}$}
\email{yuwangwei@mail.itp.ac.cn}

\affiliation{$^1$School of Physical Sciences, University of Chinese Academy of Sciences (UCAS), Beijing 100049, China}
\affiliation{$^2$CAS Key Laboratory of Theoretical Physics, Institute of Theoretical Physics, Chinese Academy of Sciences, P.O. Box 2735, Beijing 100190, China}
\affiliation{$^3$School of Fundamental Physics and Mathematical Sciences, Hangzhou Institute for Advanced Study (HIAS), University of Chinese Academy of Sciences, Hangzhou 310024, China}
\affiliation{$^4$Quantum Universe Center and School of Physics, Korea Institute for Advanced Study (KIAS), Seoul 02455, Korea}

\begin{abstract}
The increasing tension between the different local direct measurements of the Hubble expansion rate and that inferred from the cosmic microwave background observation by the $\Lambda$-cold-dark-matter model could be a smoking gun of new physics, if not caused by either observational systematics or local bias. We generalize previous investigation on the local bias from a local void by globally fitting the Pantheon sample over all parameters in the radial profile function of a local void described by an inhomogeneous but isotropic Lema\^{i}tre-Tolman-Bondi metric with a cosmological constant. Our conclusion strengthens the previous studies that the current tension on Hubble constant cannot be saved by a local void alone.
\end{abstract}

\maketitle

\section{Introduction}

The precision cosmology from the local observations of the type Ia supernovae (SNe Ia)  \cite{Riess:1998cb,Perlmutter:1998np} and the global observations of the cosmic microwave background (CMB) \cite{Smoot:1992td,Spergel:2003cb,Ade:2013sjv} has favored the dubbed $\Lambda$-cold-dark-matter ($\Lambda$CDM) model \cite{Ade:2013zuv,Ade:2015xua,Aghanim:2018eyx} as the concordance model withstanding many other data testings in the last two decades but with a notable exception for the increasing tension on the Hubble constant ($2.5\sigma$ \cite{Riess:2011yx}, $3.4\sigma$ \cite{Riess:2016jrr}, $3.7\sigma$ \cite{Riess:2018uxu}, $3.8\sigma$ \cite{Riess:2018byc}, $4.4\sigma$ \cite{Riess:2019cxk}, $4.7\sigma$ \cite{Camarena:2019moy}, and $5.3\sigma$ \cite{Wong:2019kwg}) between the local and global observations \cite{Verde:2019ivm}, which, if not caused by either systematics errors or local bias, could be the smoking gun of new physics \cite{Freedman:2017yms} beyond the $\Lambda$CDM model either from the early or late Universe \cite{Verde:2019ivm,Park:2019tyw,Knox:2019rjx}. Therefore, it is crucial to rule out the resolution from the possibility of, for example, a local void \cite{Keenan:2012gr,Keenan:2013mfa,Whitbourn:2013mwa,Whitbourn:2016irk,Boehringer:2019xmx}.

A cosmic void with its matter distribution changing in the radial coordinate could be described by the dubbed Lema\^{i}tre-Tolman-Bondi (LTB) \cite{Lemaitre:1933gd, Tolman:1934za,Bondi:1947fta} metric for an inhomogeneous but isotropic Universe we might live in locally \cite{Celerier:1999hp,Yoo:2010qy,Marra:2007pm,Brouzakis:2007zi,Kainulainen:2009sx}. However, some constraints from the detection of the secondary CMB effect like kinetic Sunyaev-Zel'dovich (kSZ) effect \cite{Sunyaev:1972eq,Sunyaev:1980nv} have ruled out a class of giant local dust void models \cite{Zhang:2010fa} (see, however, \cite{Ding:2019mmw} for a recent attempt to ease the Hubble tension but still evading the kSZ limit.). Nevertheless, the LTB model with a cosmological constant, called the $\Lambda$LTB model, still seems to be able to relieve \cite{Marra:2013rba,Wu:2017fpr,Camarena:2018nbr} or even fully resolve \cite{Tokutake:2017zqf,Hoscheit:2018nfl,Shanks:2018rka} the Hubble tension when using the galaxy survey data. For example, the luminosity density sample \cite{Lawrence:2006de,Driver:2010dp,Driver:2010zb,Lavaux:2011zu} is constructed over the redshift range $0.01<z<0.2$ for the discovery of the Keenan-Barger-Cowie (KBC) void \cite{Keenan:2013mfa} with a size of $\sim300$ Mpc and density contrast of $-30\%$. In particular, by adopting a radial profile for the matter density fraction with the Garcia-Ballido-Haugb{\o}lle (GBH) parameterization \cite{GarciaBellido:2008nz} smoothly connecting two homogeneous parts inside and outside a local underdensity,  Hoscheit and Barger \cite{Hoscheit:2018nfl} have fitted the SNe Ia data  in the redshift range $0.0233<z<0.15$ and then reduced the Hubble tension from $3.4\sigma$ to $2.75\sigma$ with the GBH parameters fixed by the KBC void configuration.

However, the data analysis of \cite{Hoscheit:2018nfl} was revised by Kenworthy, Scolnic, and Riess in \cite{Kenworthy:2019qwq} by using a larger sample of low-redshift SNe from a combined sample of the Pantheon, Foundation, and Carnegie-Supernova-Project (CSP) samples within $z<0.5$ with fully appreciating for systematic uncertainties in the SNe data, such as uncertainties in the nuisance parameters, calibration uncertainties, and possible redshift evolution of the nuisance parameters. Furthermore, Kenworthy, Scolnic, and Riess \cite{Kenworthy:2019qwq} also adopted more physically motivated boundary conditions, such as the GBH parameterization for the physical matter density profile $\rho_M(r)$ instead of the dimensionless matter density fraction $\Omega_M(r)$ used in \cite{Hoscheit:2018nfl}, and the treatment of the time since the big bang $t_B$ as a free constant parameter instead of the inappropriate choice $t_B(r)=r$ used in \cite{Hoscheit:2018nfl}. Nevertheless, the GBH profile in \cite{Kenworthy:2019qwq} was also fixed by the same KBC void configuration. The conclusion drawn from \cite{Kenworthy:2019qwq} agrees well with previous studies \cite{Wojtak:2013gda,Odderskov:2014hqa,Wu:2017fpr} that the cosmic void for the Hubble constant determination fitted by the Hubble diagram is inadequate to account for the current discrepancy of the Hubble tension. Later in \cite{Lukovic:2019ryg} the GBH parameterization was chosen for the spatial curvature in order to achieve a complete analytic determination of the cosmic time in terms of the radial coordinate suitable for fitting the void size from a top-hat profile. Consistent with Kenworthy, Scolnic, and Riess in \cite{Kenworthy:2019qwq}, the local void fitted by the low-redshift Pantheon SNe data in \cite{Lukovic:2019ryg} is also insufficient to resolve the Hubble tension, in contrast to the luminosity distance data that admit a large local void.

Although \cite{Kenworthy:2019qwq} has ruled out a local void with a sharp edge and depth $|\Delta\delta|>20\%$ in the redshift range $0.023<z<0.15$ by fitting the GBH matter profile to the combined SNe data sample (Pantheon, Foundation and CSP) within $z<0.5$, this does not automatically rule out a larger void with a shallower depth and a wider edge in a larger sample of SNe data. The void search from \cite{Lukovic:2019ryg} has used the full data of Pantheon sample but the GBH profile has been imposed on the spatial curvature term instead of the matter density fraction. 
Besides, a mild tension $(2-3\sigma)$ was found in \cite{Kazantzidis:2020tko} in the context of the $\Lambda$CDM model (instead of the $\Lambda$LTB model) between the best fit value of $\mathcal{M}\equiv M+5\log_{10}(c/H_0/\mathrm{Mpc})+25$ obtained from low-$z$ SNe data ($0.01\leq z\leq 0.2$) and the corresponding value obtained from the full Pantheon dataset. Here $M$ is the color and stretch corrected absolute magnitude of SN Ia.
Other work like \cite{Sapone:2020wwz} has also used the full data of Pantheon sample but the assumed model is the LTB model instead of the $\Lambda$LTB model. We therefore extend the analysis to the $\Lambda$LTB model by fitting the full Pantheon SNe data ranging from $0.01<z<2.3$ over all three parameters in the GBH parameterization for the matter density fraction. We confirm the previous findings that even in this general setting the local void cannot fully resolve the Hubble tension.  The rest of this paper is organized as follows: In Sec. \ref{sec:model}, we introduce the $\Lambda$LTB model with GBH profile function. In Sec. \ref{sec:data}, we use the Pantheon data to constrain the GBH void profile. Section \ref{sec:condis} is devoted to conclusion and discussions.

\section{GBH void in $\Lambda$LTB model}\label{sec:model}

\subsection{FLRW equation}

As a generalization of the usual Friedman-Lema\^{i}tre-Robertson-Walker (FLRW) metric within the framework of general relativity, the LTB metric \cite{Lemaitre:1933gd, Tolman:1934za,Bondi:1947fta} (see also\cite{Kenworthy:2019qwq}) uses the generalized scale factor $ R(r,t) $ and a curvature term $ k(r) $ to describe an inhomogeneous but isotropic void by
\begin{align}
	ds^2 = dt^2 - \frac{R'^2(r,t)}{1 - k(r)} dr^2 - R^2(r,t) d\Omega^2,
\end{align}
with  $R'(r,t) = \partial R(r,t)/\partial r$. Imposing the homogeneous condition $R(r,t) = a(t) r$ and $k(r) = k r^2$ would reduce the LTB metric into the FLRW metric with $a(t)$ acting as the usual cosmic scale factor and thus preserve the homogeneity and Copernican principle of our Universe. The corresponding Friedmann equation for the $ \Lambda $LTB model reads
\begin{equation}\label{eq:FLRW}
\begin{aligned}
& H^2(r,t)  \\ & = H_0^2(r) \bigg[\Omega_M (r) \bigg(\frac{R_0(r)}{R(r,t)}\bigg)^3+\Omega_k (r) \bigg(\frac{R_0(r)}{R(r,t)}\bigg)^2+\Omega_\Lambda (r)\bigg],
\end{aligned}
\end{equation}
where \begin{align}
H^2(r,t) \equiv \bigg(\frac{\dot{R}(r,t)}{R(r,t)}\bigg)^2
\end{align}
is the Hubble parameter, and the dot is taken for the derivative with respect to $t$. Hence $H_0(r)\equiv H(r, t_0)$ and $R_0(r) \equiv R(r,t_0)$. The Friedmann equation for the $\Lambda$LTB model at present time could also be written as $\Omega_M(r) + \Omega_k(r) + \Omega_\Lambda(r) = 1$.

\subsection{GBH profile}

\begin{figure}
	\includegraphics[width=0.45\textwidth]{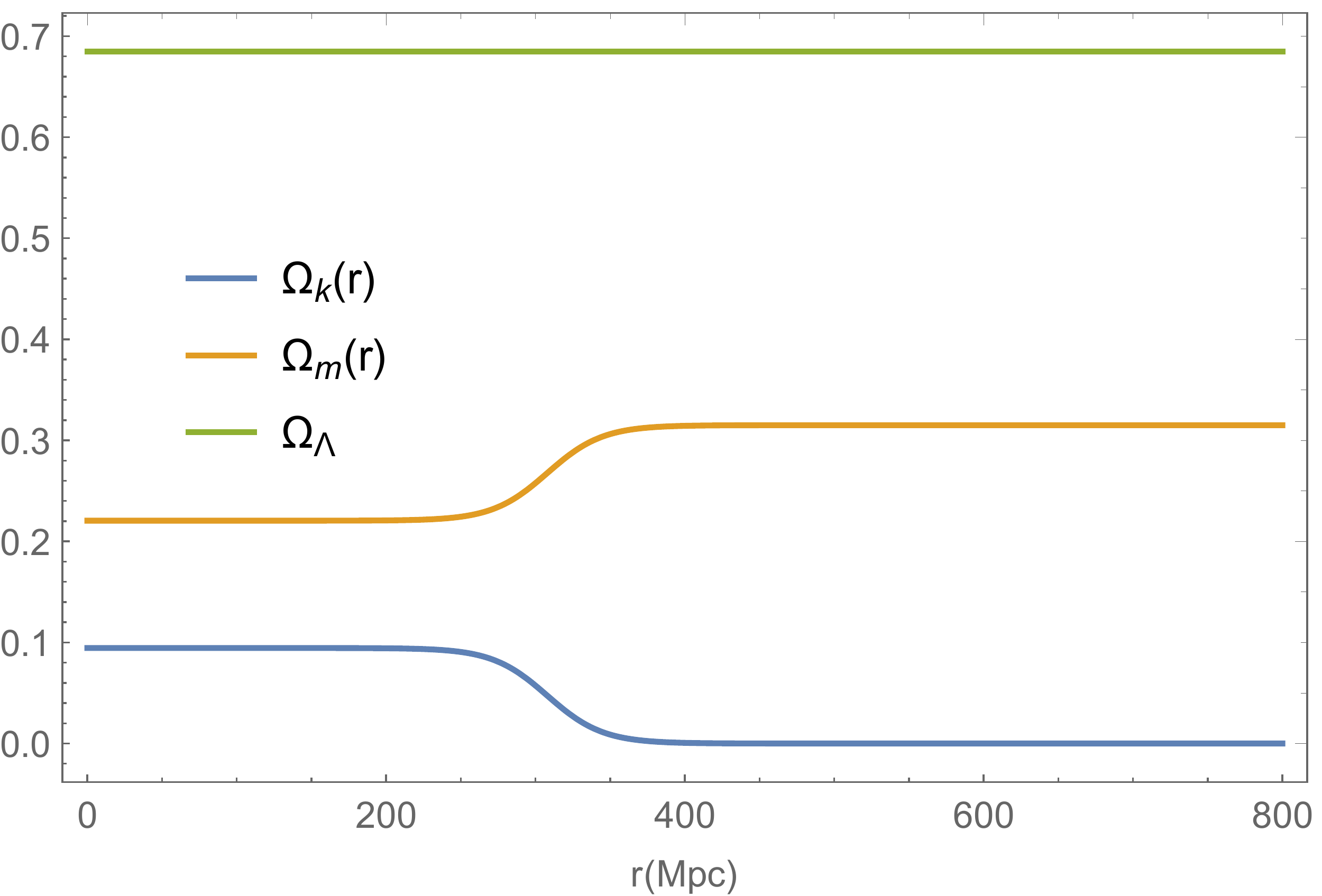}\\
	\caption{The GBH profile with KBC void parameters $\delta_V = -0.3$, $r_V = 308$ Mpc and $\Delta_r = 18.46$ Mpc.}
	\label{fig:DenCon}
\end{figure}

If the matter density parameter outside the void $ \Omega_{M,\mathrm{out}} $ is assumed to be set by the CMB data, and the matter density parameter in the void interior $ \Omega_M(r) $ exhibits explicit radial dependence, the fractional deficit could therefore be defined as 
\begin{align}
	\delta(r) \equiv \frac{\Omega_M(r) - \Omega_{M,\mathrm{out}}}{\Omega_{M,\mathrm{out}}},
\end{align}
which could be further parameterized by the dubbed GBH profile function \cite{GarciaBellido:2008nz,Keenan:2013mfa, Kenworthy:2019qwq} as 
\begin{align}
	\delta(r) = \delta_V \frac{1 - \tanh((r - r_V)/2\Delta_r)}{1 + \tanh(r_V/2\Delta_r)},
\end{align}
with $\delta_V$, $r_V$, and $\Delta_r$ characterizing the depth, radius, and transition width of the void, respectively. An illustration for the GBH profile is shown in Fig. \ref{fig:DenCon}. Besides the GBH profile function describing a Universe with \textit{two} ``homogeneous" parts linked by a smooth function, there are many other profiles like those proposed in  \cite{Vargas:2015ctw,February:2009pv,Enqvist:2006cg}. Therefore, the critical densities of matter, dark energy, and curvature could be obtained by
\begin{align}
\Omega_M(r) &= \Omega_{M,{\rm out}} (1+\delta(r)),\\
\Omega_{\Lambda}(r) &= 1 -\Omega_{ M, {\rm out}},\\
\Omega_k(r) &= 1-\Omega_M(r)-\Omega_{\Lambda}(r),
\end{align}
respectively, where $\Omega_\Lambda(r)$ is assumed to be a constant since it is not a diluted background parameter.

\subsection{Synchronous comoving gauge}

\begin{figure}
	\includegraphics[width=0.45\textwidth]{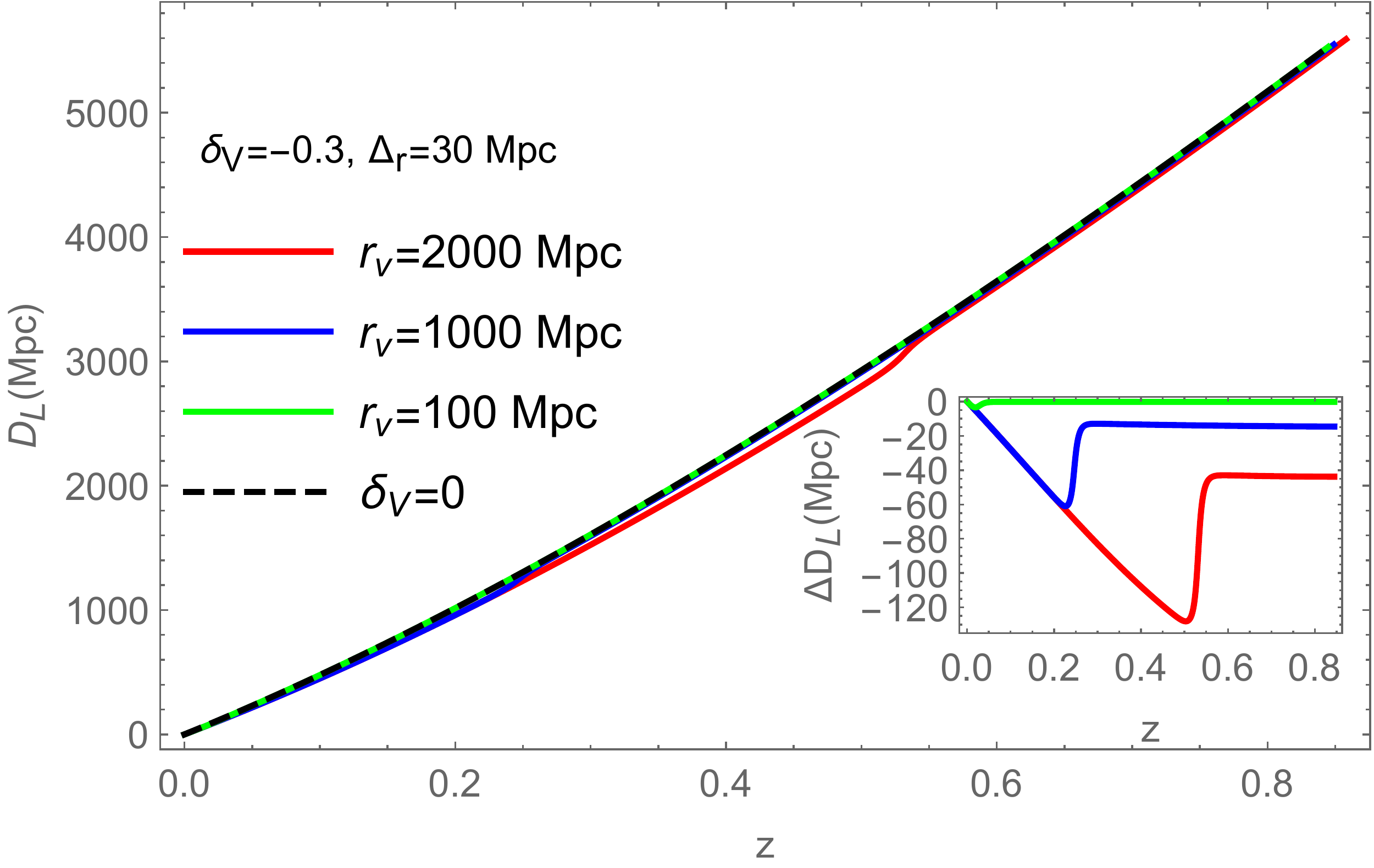} \\
	\includegraphics[width=0.45\textwidth]{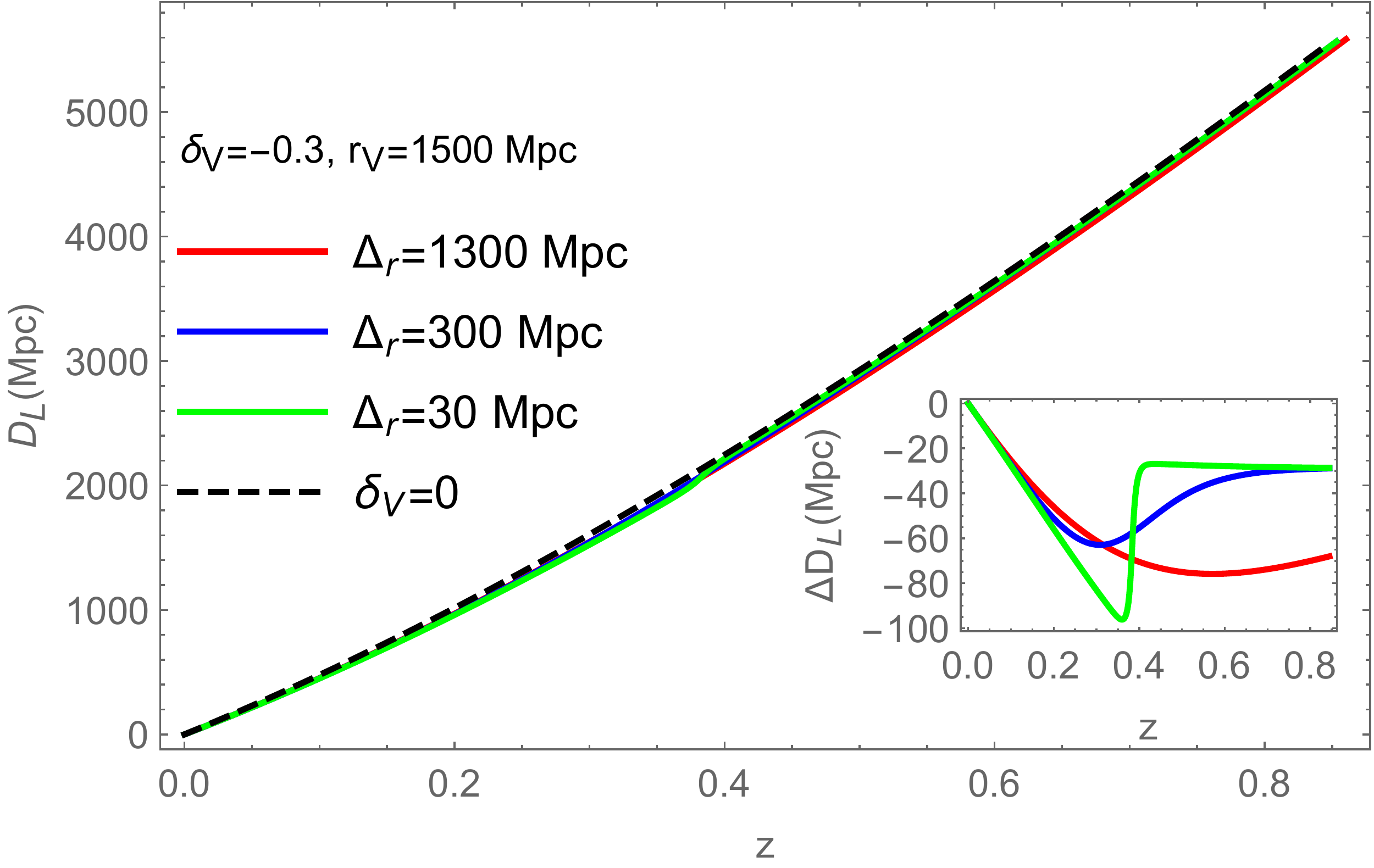} \\
	\includegraphics[width=0.45\textwidth]{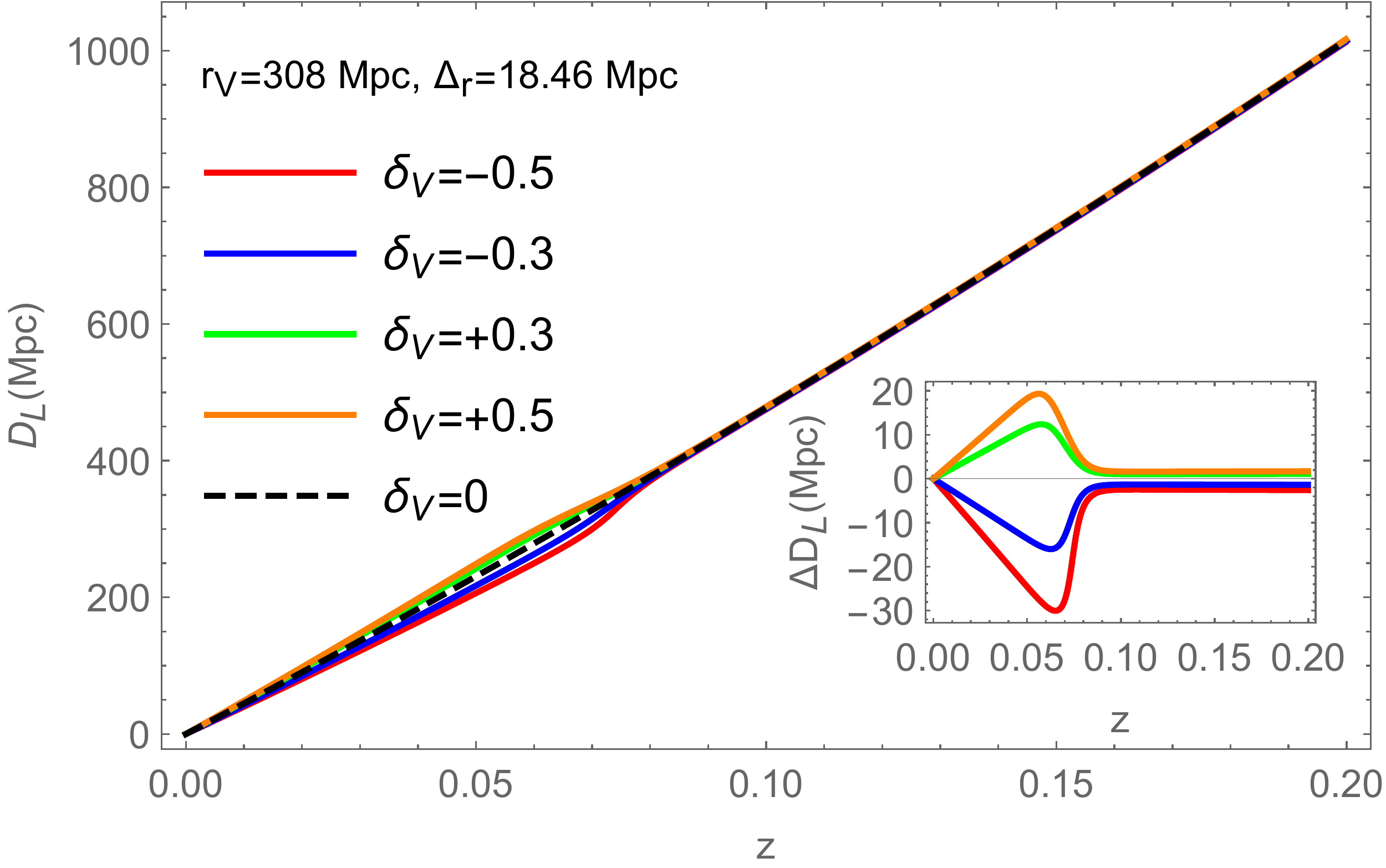}
	\caption{The luminosity distance with respect to the redshift from the $\Lambda$CDM model (black dashed curve) and $\Lambda$LTB model (solid curves) with varying $r_V$ (first panel), varying $\Delta_r$ (second panel), and varying $\delta_V$ (third panel), respectively, but with the other two parameters fixed as indicated in each panel. The subfigures present the differences of the corresponding $\Lambda$LTB models with respect to the $\Lambda$CDM model.}\label{fig:fittings}
\end{figure}

We choose the synchronous comoving gauge  $ R_0(r) = r$ for \eqref{eq:FLRW}:
\begin{equation}
\begin{aligned}
	H^2(r,t) =&  \Omega_{M,\mathrm{out}}(1+\delta (r)) H_{0}^2(r) \bigg(\frac{r}{R(r,t)}\bigg)^3 -  \Omega_{M,\mathrm{out}} \\
	 &  \delta (r) H_{0}^2(r)  \bigg(\frac{r}{R(r,t)} \bigg)^2+
	(1-\Omega_{M,\mathrm{out}})  H_{\mathrm{0}}^2(r).
\end{aligned}
\end{equation}
which, after integrated, gives rise to the age of the Universe of form
\begin{equation}\label{eq:LLTBtime}
	\begin{aligned}
t_B(r) &= \int_0^r dR R^{-1} \bigg[\Omega_M(r) H_0^2(r) \left(\frac{r}{R}\right)^3 
+ \Omega_k(r) H_0^2(r) \\
& \bigg(\frac{r}{R}\bigg)^2 + \Omega_\Lambda(r) H_0^2(r) \bigg]^{-1/2}.
	\end{aligned}
\end{equation}
For $r$ at CMB scales with $ \delta(r) = 0 $, every physical parameter is in accordance with its CMB counterpart so that we can dismiss the difference between $ t_B(r) $ and the real cosmic time $ t_0(r) $ and set the universal cosmic time assumption $t_B (r) = t_B \equiv \mathrm{const}$. \cite{Kenworthy:2019qwq}, where $t_B$ is the time at CMB scales:
\begin{align}\label{eq:FLRWtime}
t_B = \int_0^1 \frac{d a_{\mathrm{out}}(t)}{H_{\mathrm{0,out}} [\Omega_{M,\mathrm{out}} a_{\mathrm{out}}^{-1} + \Omega_{\Lambda,\mathrm{out}} a_{\mathrm{out}}^2]^{1/2}},
\end{align}
with $ a_{\rm out}(t) $ playing the role of the scale factor outside the void at CMB scales in accordance with its counterpart in FLRW metric.  Now we can combine \eqref{eq:LLTBtime} and \eqref{eq:FLRWtime} to infer $H_0(r)$ as a function of the radial coordinate $ r $. Furthermore, using the equations for null geodesics in the $ \Lambda $LTB model could lead to the redshift $z$ as a function of the radial coordinate $r$ and the cosmic time $t$ \cite{Kenworthy:2019qwq}:
\begin{equation}
\begin{aligned}
 \frac{dt}{dr} = - \frac{R'(r,t)}{\sqrt{1-k(r)}}, \\
 \frac{1}{1+z}\frac{dz}{dr} = \frac{\dot{R}'(r,t)}{\sqrt{1-k(r)}}.
\end{aligned}	
\end{equation}
which could be inverted to rewrite the cosmic time $ t $, the radial coordinate $ r $ and every other parameter within the void in terms of the redshift $z$ including, for example, the luminosity distance, 
\begin{equation}\label{eq:dL}
	d_L = (1+z)^2 R(r(z),t(z)).
\end{equation}  
To manifest the difference of using the $\Lambda$LTB model with respect to the $\Lambda$CDM model, the luminosity distance from varying one of the GBH parameters but fixing the other two GBH parameters has been presented in Fig.~\ref{fig:fittings} with solid curves compared to the $\Lambda$CDM result shown as a black dashed curve. 

\section{Data analysis and cosmological constraint}\label{sec:data}

\begin{figure*}
\includegraphics[width=0.32\textwidth]{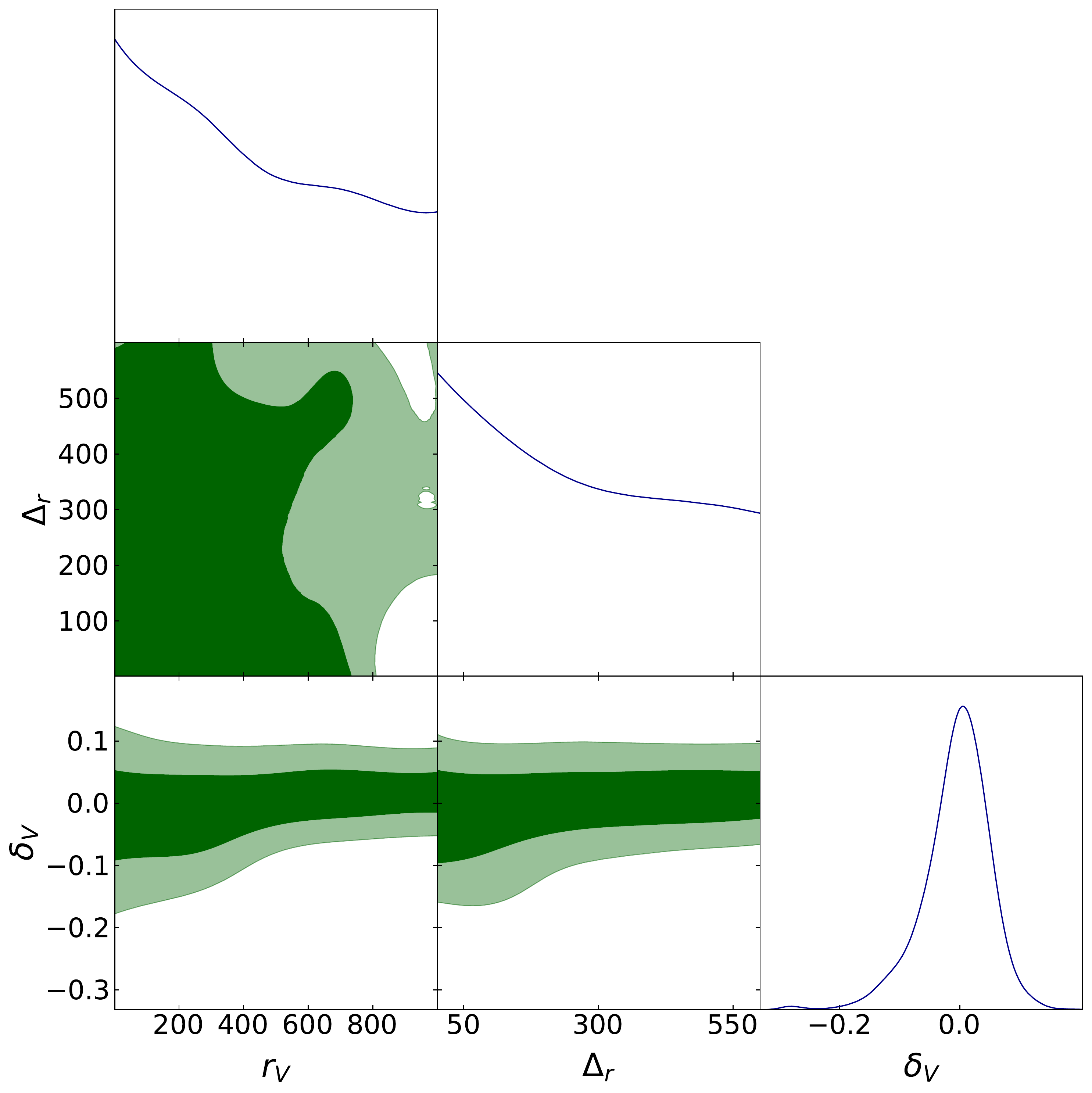}
\includegraphics[width=0.32\textwidth]{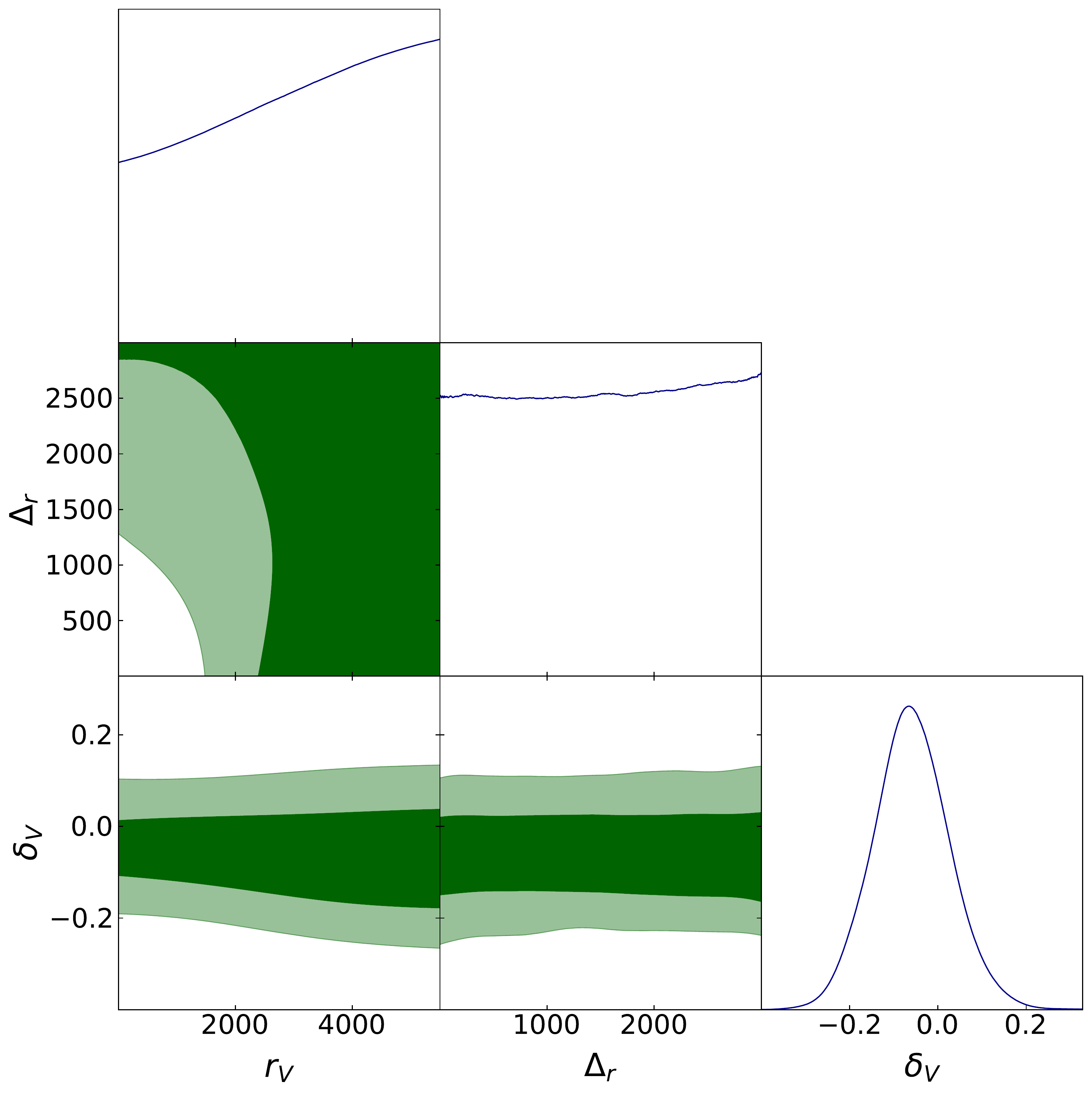}
\includegraphics[width=0.32\textwidth]{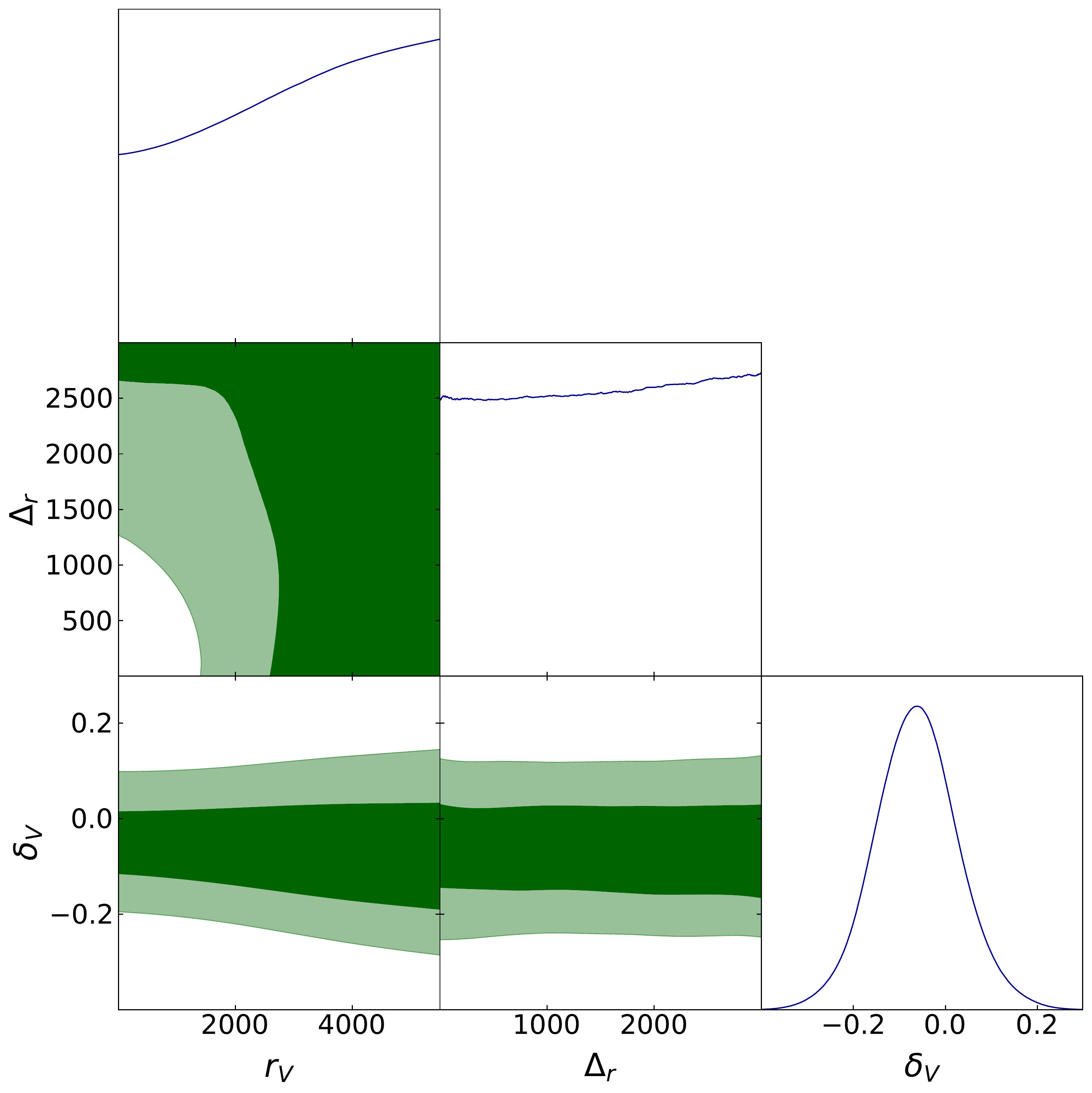}\\
\caption{Posterior constraints on $ \delta_V $, $ \Delta_r $ and $ r_V $ from searching the void within redshift $z\leq0.2$, and $z\leq1$, $z\leq2$, respectively. The contours describe the 68\% and 95\% confidence limits.}\label{fig:MCMC}
\end{figure*}

\subsection{Data analysis}

The data we use for analysis in this paper consist of 1048 SNe Ia ranging from $ 0.01 < z < 2.3 $, which is summarized in the Pantheon sample \cite{Scolnic:2017caz} combining the following datasets with consistent photometry:
\begin{itemize}
\item[(i)]  \textit{Low}-$z$ \textit{data}.---$ 0.01 < z <0.1 $ from CfA1-4 \cite{Riess:1998dv,Jha:2005jg,Hicken:2009df,Hicken:2009dk,Hicken:2012zr} and CSP \cite{Contreras:2009nt,Folatelli:2009nm,Stritzinger:2011qd};
\item[(ii)] \textit{Intermediate}-$z$ \textit{data}.---$ 0.03 \lesssim z \lesssim 0.68 $ from PS1 \cite{Rest:2013mwz,Scolnic:2013efb}, $ 0.1 \lesssim z \lesssim 0.4 $ from SDSS \cite{Frieman:2007mr,Kessler:2009ys,Sako:2014qmj}, and $ 0.3 \lesssim z \lesssim 1.1 $ from SNLS \cite{Conley:2011ku,Sullivan:2011kv} ;
\item[(iii)] \textit{High}-$z$ \textit{data}.---$ z>1.0 $ from SCP \cite{Suzuki:2011hu}, GOODS \cite{Riess:2004nr,Riess:2006fw} and CANDELS/CLASH \cite{Rodney:2014twa,Graur:2013msa,Riess:2017lxs}.
\end{itemize}
The above data will be tested for our $\Lambda$LTB model with GBH void profile function in the redshift-distance relation \eqref{eq:dL}, where the luminosity distance is measured as usual by the distance modulus,
\begin{equation}
	\mu = m_B^0 - M_B^0 = 5{\rm log}_{\rm 10}\left(\frac{d_L(z)}{\rm Mpc}\right) + 25,
\end{equation}
where $ m_B^0 $ is the corrected peak magnitude of a SN and $ M_B^0 $ is the absolute magnitude of a fiducial counterpart. It is worth noting that, for Pantheon data, it is necessary to consider the heliocentric redshift $z_{\mathrm{Hel}}$ in
\begin{equation}
\begin{aligned}
\mu & = m_B^0 - M_B^0  \\ 
	& = 5{\rm log}_{\rm 10}\left(\frac{d_L(z_{\mathrm{CMB}})}{\rm Mpc}\right) + 5 \log_{10}\left(\frac{1+z_{\mathrm{Hel}}}{1+z_{\mathrm{CMB}}}\right) + 25.
\end{aligned}
\end{equation}

The test we adopt for data analysis is the usual $ \chi^2 $ test~\cite{Scolnic:2017caz}:
\begin{equation}
	\chi^2 = \Delta \boldsymbol{\mu}^T\cdot \textbf{C}^{-1} \cdot \Delta \boldsymbol{\mu},
\end{equation}
where  $ \Delta \boldsymbol{\mu} = \boldsymbol{\mu} - \boldsymbol{\mu}_{\rm \Lambda LTB} $ and $ \textbf{C} $ is the covariance matrix consisting of
\begin{equation}
	\textbf{C} = \textbf{D}_{\mathrm{stat}} + \textbf{C}_{\mathrm{sys}}.
\end{equation} 
$\textbf{D}_{\mathrm{stat}}$ has only  diagonal components containing the total distance errors associated with each SN, which include photometric error, mass step correction, distance bias correction, peculiar velocity uncertainty, redshift measurement uncertainty in quadrature, stochastic gravitational lensing and intrinsic scatter. $\textbf{C}_{\mathrm{sys}}$ is the systematic covariance. The method we use for data analysis is the usual Markov chain Monte Carlo (MCMC) sampling \cite{Lewis:2002ah} to scan all three  parameters ($ r_V, \Delta_r $ and $ \delta_V $) of the GBH profile in the $ \Lambda $LTB model  with respect to the Pantheon data \cite{Scolnic:2017caz} and the $\Lambda$CDM model calibrated by the Planck 2018 results $ \Omega_{M,{\rm out}}=0.315 \pm 0.007 $ and $H_{0,\mathrm{out}} = (67.4 \pm 0.5)$ km/s/Mpc \cite{Aghanim:2018eyx} outside the void in order to match the CMB observations at large scales. Since we have no \textit{a priori} setup for the void we want to identify by the Pantheon data alone, we therefore choose the following flat prior for all three parameters in the GBH parameterization with three illustrative redshift bins:
\begin{itemize}
\item[(i)] $z<0.2$.---$r_V\in[1,1000]$ Mpc,  $\Delta_r\in[1,600]$ Mpc, $\delta_V\in[-0.5,0.5]$;
\item[(ii)] $z<1$.---$r_V\in[1,5500]$ Mpc,  $\Delta_r\in[1,3000]$ Mpc, $\delta_V\in[-0.4,0.4]$;
\item[(iii)] $z<2$.---$r_V\in[1,5500]$ Mpc,  $\Delta_r\in[1,3000]$ Mpc, $\delta_V\in[-0.4,0.4]$.
\end{itemize}
Note that the shortest distance to a supernova in our data sample is $\sim$ 45 Mpc (CMB frame) and the local void search is limited within redshift $ z\le 2 $ since most of SNe data of Pantheon sample are within redshift $ z \le 2 $ (1047 of 1048 in total) so that the $ \Lambda $LTB metric is reduced to the FLRW metric outside $z>2$ with the homogeneous condition and $ k(r) = kr^2 $ automatically preserved.

\subsection{Cosmological constraints}

\begin{table*}[!ht]
\caption{The cosmological constraints from the global fitting for the total Pantheon sample cut at different redshift ranges (first column) with corresponding number of SNe (second column) on the best-fit value (third column), mean  value (fourth column), and standard deviation (fifth column) of the void depth $\delta_V$, the best-fit value of the void radius $r_V$  (sixth column), the best-fit value of the void transition width $\Delta_r$ (seventh column), the reduced $\chi^2$ for the best-fit $\Lambda$LTB model  (eighth column), the reduced $\chi^2$ for the $\Lambda$CDM model (ninth column), the comparison to $\Lambda$CDM model using the relative AIC values (tenth column), and the comparison to the $\Lambda$CDM model using the relative BIC values (11th column).}\label{tab:multiple}
\begin{tabular}{c|c|c|c|c|c|c|c|c|c|c}
\hline
\hline
SNe  
& SNe 
& \multicolumn{3}{|c|}{$\delta_V$} 
& $r_V$ & $\Delta_r$
& \multirow{2}*{$\frac{\chi_{\Lambda\mathrm{LTB}}^2}{\mathrm{d.o.f.}}$}
& \multirow{2}*{$\frac{\chi_{\Lambda\mathrm{CDM}}^2}{\mathrm{d.o.f.}}$}
& \multirow{2}*{$\Delta\mathrm{AIC}$}
& \multirow{2}*{$\Delta\mathrm{BIC}$}
\\
\cline{3-5}
range  & number & Best Fit    & Mean       & Std-dev & [Mpc] & [Mpc] & & & \\
\hline
$z\leq0.2$  & 411 & $-9.1\%$  & $-5.6\%$ & $5.8\%$ & 334  & 1.71 & 0.988   & $0.988$ & $2.61$ & $17.47$ \\
$z\leq1.0$   & 1025 & $-5.5\%$ & $-5.9\%$ & $8.7\%$ & 224  & 31.03& 0.989 & $0.988$ & $4.21$ & $19.07$ \\
$z\leq2.0$  & 1047 & $-6.2\%$ & $-6.2\%$ & $9.1\%$ & 1204 & 2.02 & 0.987   & $0.988$ & $1.90$ & $16.77$ \\
\hline
\hline
\end{tabular}
\end{table*}

\begin{table*}[!ht]
	\caption{The cosmological constraints from fitting over $\delta_V$ alone with the other two GBH parameters fixed at the KBC void parameters ($r_V=308$ Mpc, $\Delta_r=18.46$ Mpc) for the total Pantheon sample cut at different redshift ranges (first column) with corresponding number of SNe (second column) on the best-fit value (third column), mean  value (fourth column), and standard deviation (fifth column) of the void depth $\delta_V$, the reduced $\chi^2$ for the best-fit $\Lambda$LTB model (sixth column), the reduced $\chi^2$ for the $\Lambda$CDM model (seventh column), the comparison to $\Lambda$CDM model using the relative AIC values (eighth column), and the comparison to $\Lambda$CDM model using the relative BIC values (ninth column).}\label{tab:single}
	\begin{tabular}{c|c|c|c|c|c|c|c|c}
		\hline
		\hline
		SNe  
		& SNe 
		& \multicolumn{3}{|c|}{$\delta_V$} 
		& \multirow{2}*{$\frac{\chi_{\Lambda\mathrm{LTB}}^2}{\mathrm{d.o.f.}}$}
		& \multirow{2}*{$\frac{\chi_{\Lambda\mathrm{CDM}}^2}{\mathrm{d.o.f.}}$}
		& \multirow{2}*{$\Delta\mathrm{AIC}$}
		& \multirow{2}*{$\Delta\mathrm{BIC}$}
		\\
		\cline{3-5}
		range  & number & Best Fit    & Mean       & Std-dev & & & \\
		\hline
		$z\leq0.2$  & 411 & $-7.4\%$  & $-7.1\%$ & $4.8\%$ & $0.987$ & $0.988$ & $-0.40$ & $4.55$  \\
		$z\leq1.0$   & 1025 &  $-8.4\%$ & $-8.2\%$ & $4.5\%$ & $0.986$ & $0.988$ & $-1.09$ & $3.87$  \\
		$z\leq2.0$  & 1047 & $-8.5\%$ & $-8.2\%$ & $4.7\%$  & $0.986$ & $0.988$ & $-1.12$ & $3.84$  \\
		\hline
		\hline
	\end{tabular}
\end{table*}

The cosmological constraints are shown in Fig. \ref{fig:MCMC} and summarized in Table \ref{tab:multiple}, which are marginally consistent with the case of no local void $\delta_V = 0$ and there is seemingly no constraint on $ r_V $ and $ \Delta_r $ simply because there is no difference between the inside and outside of the void in the case of  $\delta_V=0$. This is not surprising since $ r_V $ and $ \Delta_r $ can only be strongly constrained when there is a strong preference for a large $\delta_V$. On the other hand, if we want to search for any local void in the full data of Pantheon sample, then we have to allow for a nonzero $\delta_V$, and there is no reason to fix $r_V$ and $\Delta_r$ anymore, nor do we have any prior value for fixing $r_V$ and $\Delta_r$; hence, we fit all three GBH parameters in this paper. Since at the limit of $\delta_V \to 0$ the void model reduces to the $\Lambda$CDM model and $R/r$ is a function of $t$ alone, namely the scale factor in $ \Lambda $CDM model,  the  kSZ effect \cite{Sunyaev:1972eq,Sunyaev:1980nv}, the Rees-Sciama effect \cite{Rees:1968zza, Granett:2008xb, Masina:2008zv} and baryon acoustic oscillations would be in accordance with the $ \Lambda $CDM model.

\begin{figure}
\includegraphics[width=0.45\textwidth]{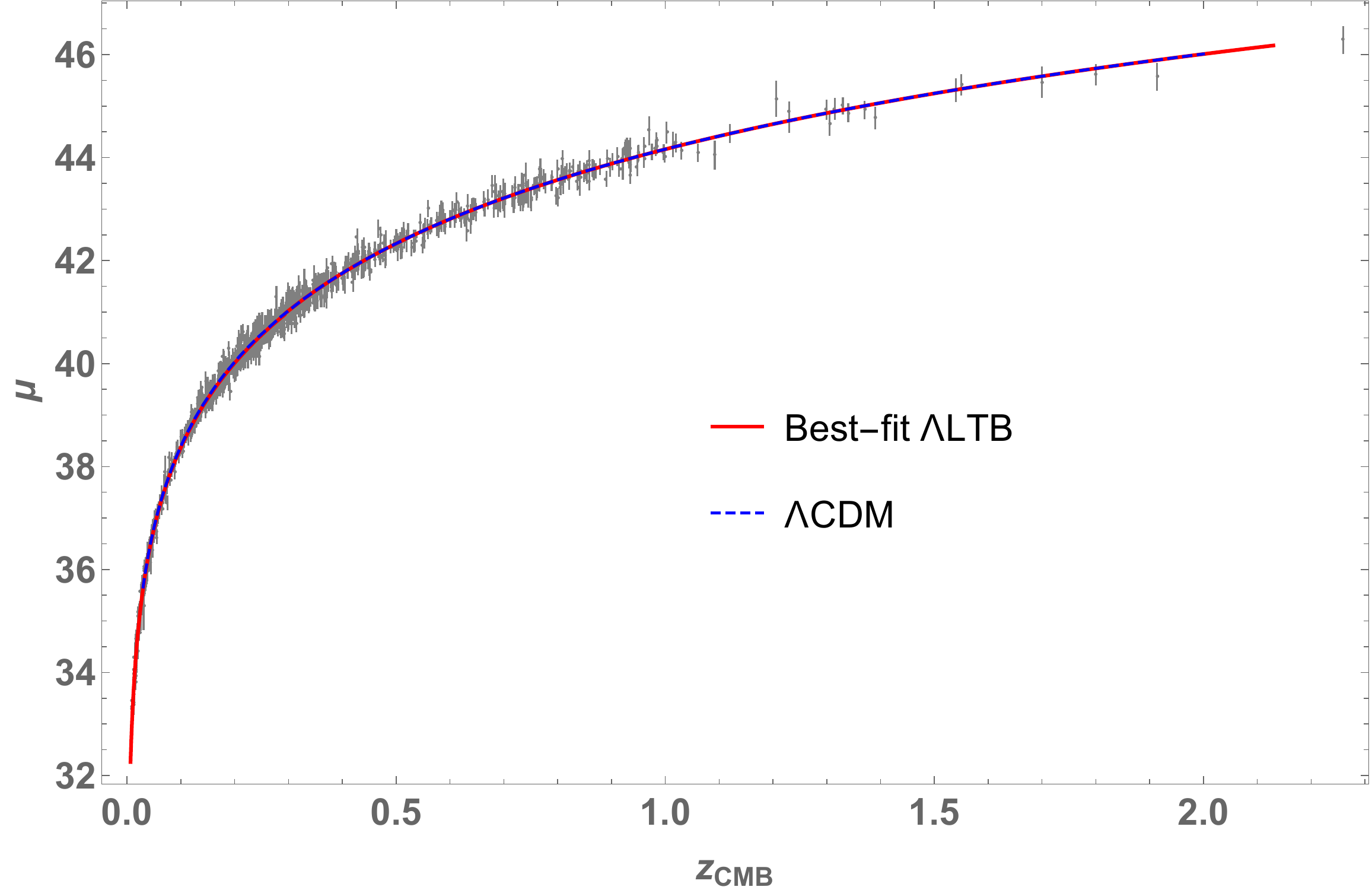}
\caption{Comparison of the distance modulus between the $\Lambda$LTB model (red solid line) with the best-fitting values constrained from SNe data within  $z\le 2$ and the $\Lambda$CDM model (blue dashed line) with values from Planck 2018 constraints. The Pantheon data are shown in gray.}
\label{fig:comparison}
\end{figure}

Therefore, our cosmological constraints  are consistent with the $ \Lambda $CDM model instead of the $\Lambda$LTB model with large $\delta_V$ in GBH parameterization for the void profile within $z \le 2$. At the very least, the $ \Lambda $CDM model cannot be distinguished from the $\Lambda$LTB model with small $\delta_V$ constrained by the SNe data alone as shown in Fig. \ref{fig:comparison}. Even if we could live in a void, the radial profile change is too insignificant to modify the concordance model.

Besides the $\chi^2$ test, we also provide with the Akaike information criterion (AIC) \cite{akaike1974new} and Bayesian information criterion (BIC) tests \cite{Schwarz:1978tpv}, which are defined by $\mathrm{AIC}=2k-2\ln \hat{L}$ and $\mathrm{BIC}=k\ln n-2\ln\hat{L}$, respectively, where $k$ is the number of parameters for a given model, $n$ is the number of data points, and $\hat{L}$ is  the maximized value of the likelihood function. 
 A smaller AIC or BIC value means a better fitting for the given model. For model selection \cite{Liddle:2007fy}, we use the relative AIC or BIC value of the $\Lambda$LTB model with respect to the $\Lambda$CDM model. Therefore, a larger relative AIC or BIC value means that the $\Lambda$LTB model is less preferred compared to the $\Lambda$CDM model, and a more negative relative AIC or BIC value means that the $\Lambda$LTB model is more favorable. 
As seen from the last two columns of Table \ref{tab:multiple}, there is no strong preference between the $\Lambda$LTB and $\Lambda$CDM models according to the relative AIC values $\Delta\mathrm{AIC}$. However, all the relative BIC values $\Delta\mathrm{BIC}$ are larger than 10, which strongly disfavor the $\Lambda$LTB model over the $\Lambda$CDM model. To see how much improvement we can make from fitting all GBH parameters compared to a single parameter ($\delta_V$) fitting, we have also fitted the Pantheon SNe data of different redshift ranges to the depth of cosmic void $\delta_V$ alone while fixing the other two GBH parameters at the KBC void parameters ($r_V=308$ Mpc and $\Delta_r=18.46$ Mpc). The relative AIC and BIC values of the $\Lambda$LTB model with respect to the $\Lambda$CDM model are summarized in the last two columns of Table \ref{tab:single}, which admit no preference for the $\Lambda$LTB model over $\Lambda$CDM model. Therefore, global fitting of all GBH parameters could lead to more dramatic disfavor of the $\Lambda$LTB model over the $\Lambda$CDM model.

\begin{figure}
\includegraphics[width=0.45\textwidth]{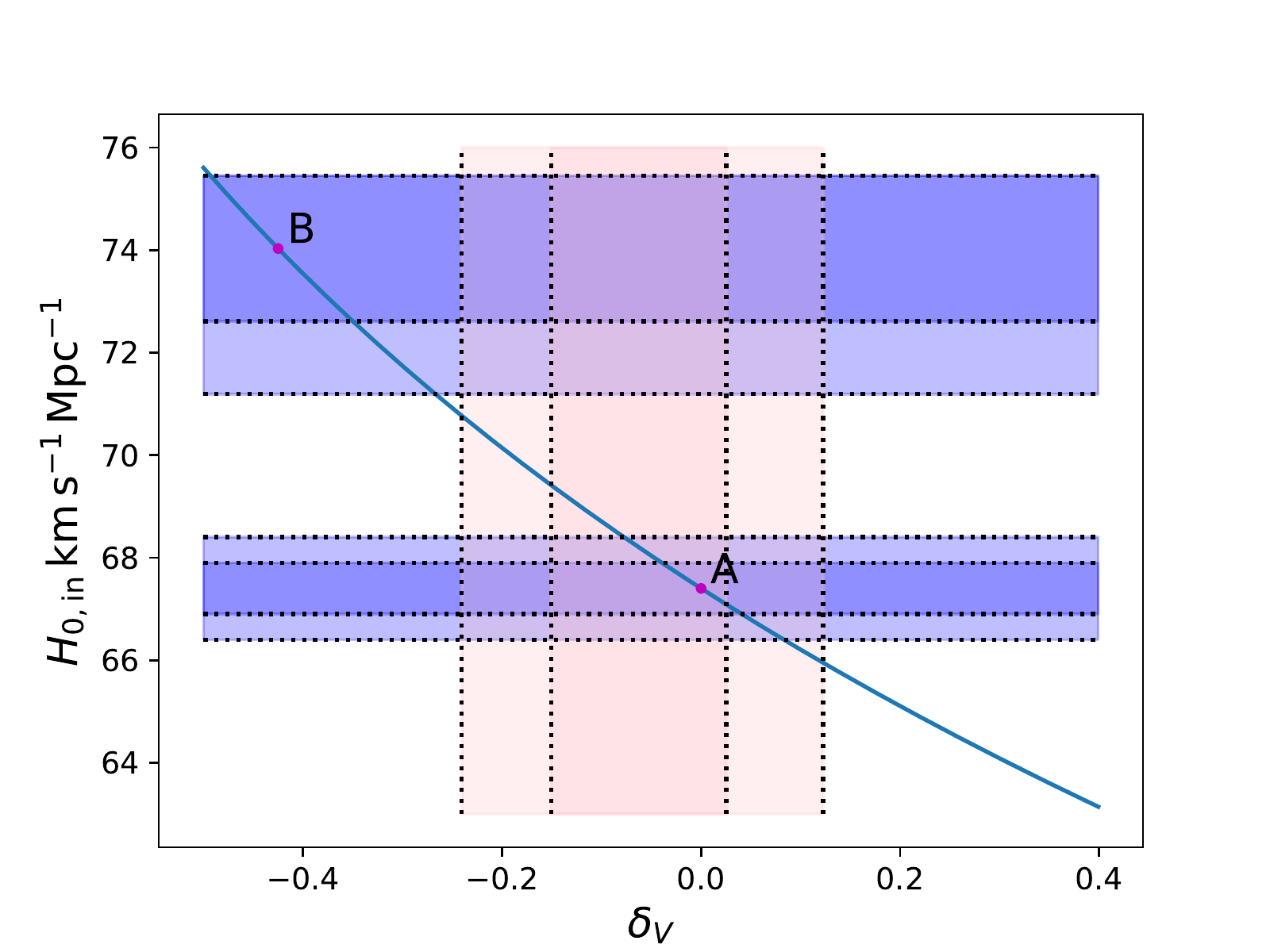}
\caption{The required $\delta_V$ for given $H_0$ inside a void described by our $\Lambda$LTB model, where ``A'' and ``B'' correspond to the Hubble constants from the Planck 2018 constraint and local measurement \cite{Riess:2019cxk}, respectively. The shaded regions are for $2\sigma$ uncertainty regions.}\label{fig:H0in}
\end{figure}

To see how large $\delta_V$ is needed for resolving the Hubble tension, we depict the Hubble constant in local void $H_{\rm 0,in}$ as a function of $ \delta_V $ assuming the Hubble constant at the CMB scales as $ H_{\rm 0, out} $. It is easy to see that $ \delta_V $ should be less than $ -30\% $ to moderately resolve the Hubble tension as shown in Fig. \ref{fig:H0in}, which is beyond the uncertainty region of  $ \delta_V $ even if all three GBH parameters are used in the data fitting.

\section{Conclusion and discussions}\label{sec:condis}

In this paper, we have constrained the $ \Lambda $LTB model with GBH parameterization for the local void profile from the MCMC sampling in fitting the  Pantheon data within $ z \le 2 $. Our fitting results are consistent with the $ \Lambda $CDM model and have no significant preference for a large local void to fully resolve the Hubble tension. However, the current analysis still merits further improvements to settle down the void issue when addressing the Hubble tension:

First, we have made a compromised choice for our GBH parameterization on the dimensionless matter density fraction instead of the physical matter density profile, the latter of which is quite challenging to implement in the numerical global fitting. The previous studies \cite{Kenworthy:2019qwq,Lukovic:2019ryg} evade this difficulty by either fixing the GBH parameters as the KBC void parameters  \cite{Kenworthy:2019qwq} or parameterizing the spatial curvature in terms of the GBH profile with top-hat shape \cite{Lukovic:2019ryg}. A GBH parameterization for the physical matter density profile in the global fitting would lead to the most general conclusion reserved for future study.

Second, we have not explored the global fitting constraint for the luminosity distance data from the galaxy survey that usually leads to the discovery of a large local void. It is unclear whether there is some unidentified systematic in the SNe data or galaxy survey data to reconcile their apparent disagreement when fitting to a local void.

\begin{acknowledgments}
We thank Wei-Ming Dai, Chang Liu, De-Yu Wang, Wen-Hong Ruan, Jing Liu and Yong Zhou for the delightful discussions and help with code. 
This work is supported in part by the National Natural Science Foundation of China Grants No. 11690021, No. 11690022, No. 11851302, No. 11821505, and No. 11947302, in part by the Strategic Priority Research Program of the Chinese Academy of Sciences Grant No. XDB23030100 and No. XDA15020701; the Key Research Program of the CAS Grant No. XDPB15; and by Key Research Program of Frontier Sciences, CAS. 
\end{acknowledgments}

\bibliographystyle{utphys}
\bibliography{ref}

\end{document}